\documentclass[]{spie}  

\usepackage{amsmath,amsfonts,amssymb}
\usepackage{graphicx}
\graphicspath{{figures/}}
\usepackage[colorlinks=true, allcolors=blue]{hyperref}
\usepackage{caption}
\usepackage{subcaption}

\title{The Bright Pyramid Wavefront Sensor}

\author[a,b]{Benjamin L. Gerard}
\author[c,d]{Vincent Chambouleyron}
\author[b,a]{Rebecca Jensen-Clem}
\author[c,d]{Jean-Fran\c{c}ois Sauvage}

\affil[a]{University of California Observatories, CA, USA}
\affil[b]{University of California Santa Cruz, CA, USA}
\affil[c]{Aix Marseille Univ, CNRS, CNES, LAM, Marseille, France}
\affil[d]{DOTA, ONERA, Universit{\'e} Paris Saclay, F-91123 Palaiseau, France
}

\authorinfo{Send correspondence to Benjamin L. Gerard: blgerard@ucsc.edu}

\begin{document} 
\maketitle

\begin{abstract}
Extreme adaptive optics (AO) is crucial for enabling the contrasts needed for ground-based high contrast imaging instruments to detect exoplanets. Pushing exoplanet imaging detection sensitivities towards lower mass, closer separations, and older planets will require upgrading AO wavefront sensors (WFSs) to be more efficient. In particular, future WFS designs will aim to improve a WFS's measurement error (i.e., the wavefront level at which photon noise, detector noise, and/or sky background limits a WFS measurement) and linearity (i.e., the wavefront level, in the absence of photon noise, aliasing, and servo lag, at which an AO loop can close and the corresponding closed-loop residual level). We present one such design here called the bright pyramid WFS (bPWFS), which improves both the linearity and measurement errors as compared to the non-modulated pyramid WFS (PWFS). The bPWFS is a unique design that, unlike other WFSs, doesn't sacrifice measurement error for linearity, potentially enabling this WFS to (a) close the AO loop on open loop turbulence utilising a tip/tilt modulation mirror (i.e., a modulated bPWFS; analogous to the procedure used for the regular modulated PWFS), and (b) reach deeper closed-loop residual wavefront levels (i.e., improving both linearity and measurement error) compared to the regular non-modulated PWFS. The latter approach could be particularly beneficial to enable improved AO performance using the bWFS as a second stage AO WFS. In this paper we will present an AO error budget analysis of the non-modulated bPWFS as well as supporting AO testbed results from the Marseille Astrophysics Laboratory. 
\end{abstract}

\keywords{Adaptive Optics, Wavefront Sensing}

\section{INTRODUCTION}
\label{sec:intro}  
Astronomical adaptive optics (AO) systems have evolved in the past three decades from providing barely diffraction-limited wavefronts at near infrared science wavelengths\cite{altair} to ``extreme'' Strehl ratios up to 90\%\cite{gpi_ao,sphere_ao}, as a result now enabling new exoplanet imaging science\cite{51eri}. Because of such progress and developments, this Extreme AO (ExAO) field is now limited by AO error budget components that were previously negligible with past-generation non-ExAO systems, which in turn limits the mass, separations, and ages of self-luminous exoplanets that ExAO-based exoplanet imaging instruments can detect and characterize\cite{gpies}. In particular, the Pyramid wavefront sensor (PWFS)\cite{pwfs} has revolutionized the field of Fourier-based AO wavefront sensing, enabling significant gains in wavefront measurement errors due to photon noise and aliasing, but at the cost of increased WFS non-linearities. A tip/tilt modulator mirror in the WFS path of an AO system decreases PWFS non-linearities, but this modulation comes with a trade-off of decreased efficiency to other measurement error terms\cite{carlos,fauvarque}. 

More recently it has been generalized that non-modulated Fourier-based WFSs---which includes the PWFS, Zernike WFS\cite{zernike, vincent_zwfs}, iQuad WFS\cite{iquad}, and more generally any WFS involving a phase mask in the upstream focal plane of a pupil imager---can be designed either (1) to be more linear (i.e., how well a WFS can reconstruct a given input wavefront, which is itself a function of input aberration strength) but at the cost of degrading measurement errors (i.e., the wavefront level at which photon noise, detector noise, and/or sky background limits a WFS measurement) or (2) to enable improved measurement errors but at the cost of degrading linearity\cite{fauvarque}. However, in this paper we propose two new Fourier-based WFSs called the bright and dark PWFS (bPWFS and dPWFS, respectively), modifying the PWFS design to enable---for the bPWFS---both improved measurement errors and linearity. In \S\ref{sec: sims} we present simulations of this concept and a corresponding error budget analysis compared to the regular PWFS. In \S\ref{sec: lab} we present preliminary validation of these bPWFS and dPWFS concepts from the AO testbed at the Marseille Astrophysics Laboratory\cite{loops_testbed}. We discuss further implications of these results and future work in \S\ref{sec: discussion}, and we conclude in \S\ref{sec: conclusion}.

\section{SIMULATIONS}
\label{sec: sims}
\subsection{Concept}
\label{sec: concept}
\begin{figure}[!h]
    \centering
    \begin{subfigure}[b]{0.5\textwidth}
        \includegraphics[width=\textwidth]{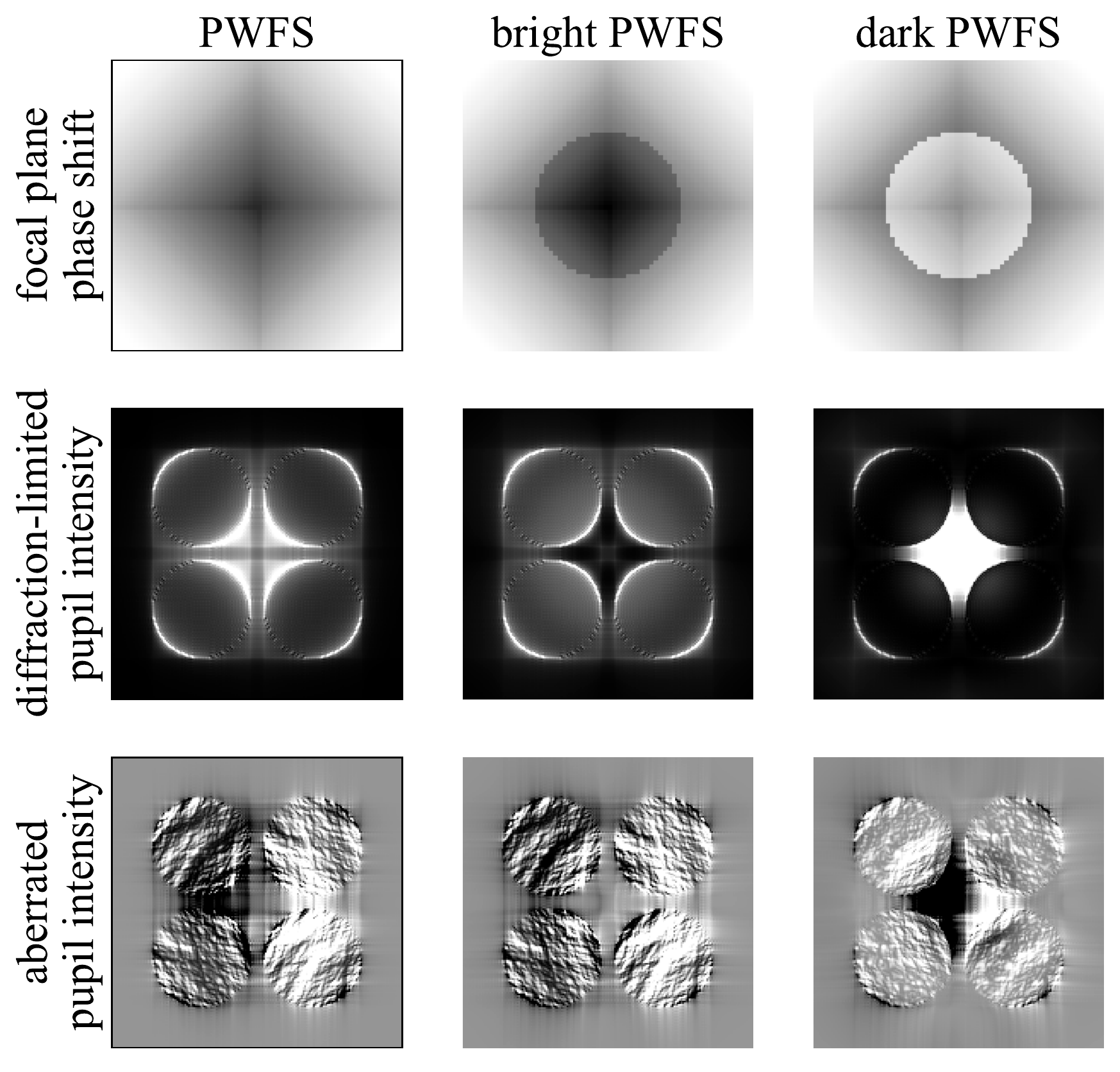}
        \caption{}
    \end{subfigure}
    \hfill
    \begin{subfigure}[b]{0.45\textwidth}
        \includegraphics[width=\textwidth]{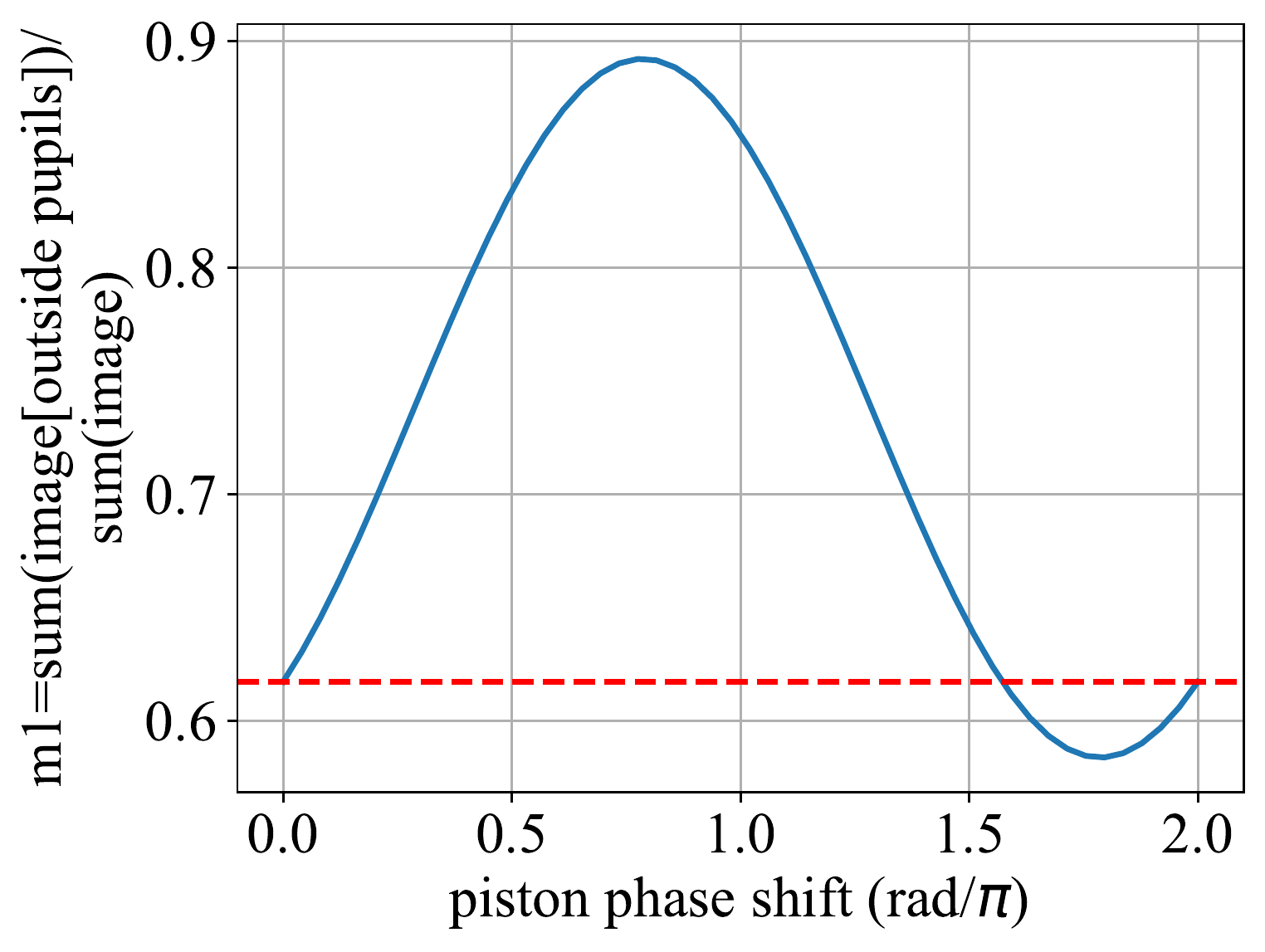}
        \caption{}
    \end{subfigure}
    \caption{(a) Illustration of the bright and dark PWFS concepts. The columns within each row are shown on the same linear contrast scale. The first row is displayed between -1 and +1 $\lambda/D$ in both x and y directions. The bright PWFS applies an added piston phase shift to the regular PWFS tilt angles of 1.7$\pi$ radians (in this case showing -0.3 rad for display purposes) within a 0.5 $\lambda/D$ radius from the pyramid apex/optical axis. The dark PWFS applies the same concept but instead with a piston phase value of 0.8$\pi$ radians. The bottom row displays the differential pupil intensity images between (1) propagated images for a 100 nm rms (at $\lambda=1.6\mu$m; normalized over 32 cycles/pupil), -2 power law phase screen realization (i.e., the same realization is used for all columns) and (2) the respective diffraction-limited images in the middle row. (b) Illustration of the diffractive nature of an additive variable piston phase shift to the PWFS tilt angles (for a diffraction-limited input wavefront). The y-axis is a metric, m1, determining how much flux is distributed beyond the four pupil footprints (i.e., in the absence of diffraction) relative total flux in the image: the lower the y axis value, the more relative flux is diffracted within the four pupil footprints. The dotted red line shows the value of m1 at piston phase shift of 0=2$\pi$ radians. When the curve is below the red dashed line, more light is diffracted within the pupil footprint than the regular PWFS; when the curve is instead above the red dashed line (which is the case for the majority of piston phases), more light light is diffracted outside of the pupil footprints compared to the regular PWFS.}
    \label{fig: bpwfs}
\end{figure}

\begin{figure}[!h]
    \centering
    \includegraphics[width=0.7\textwidth]{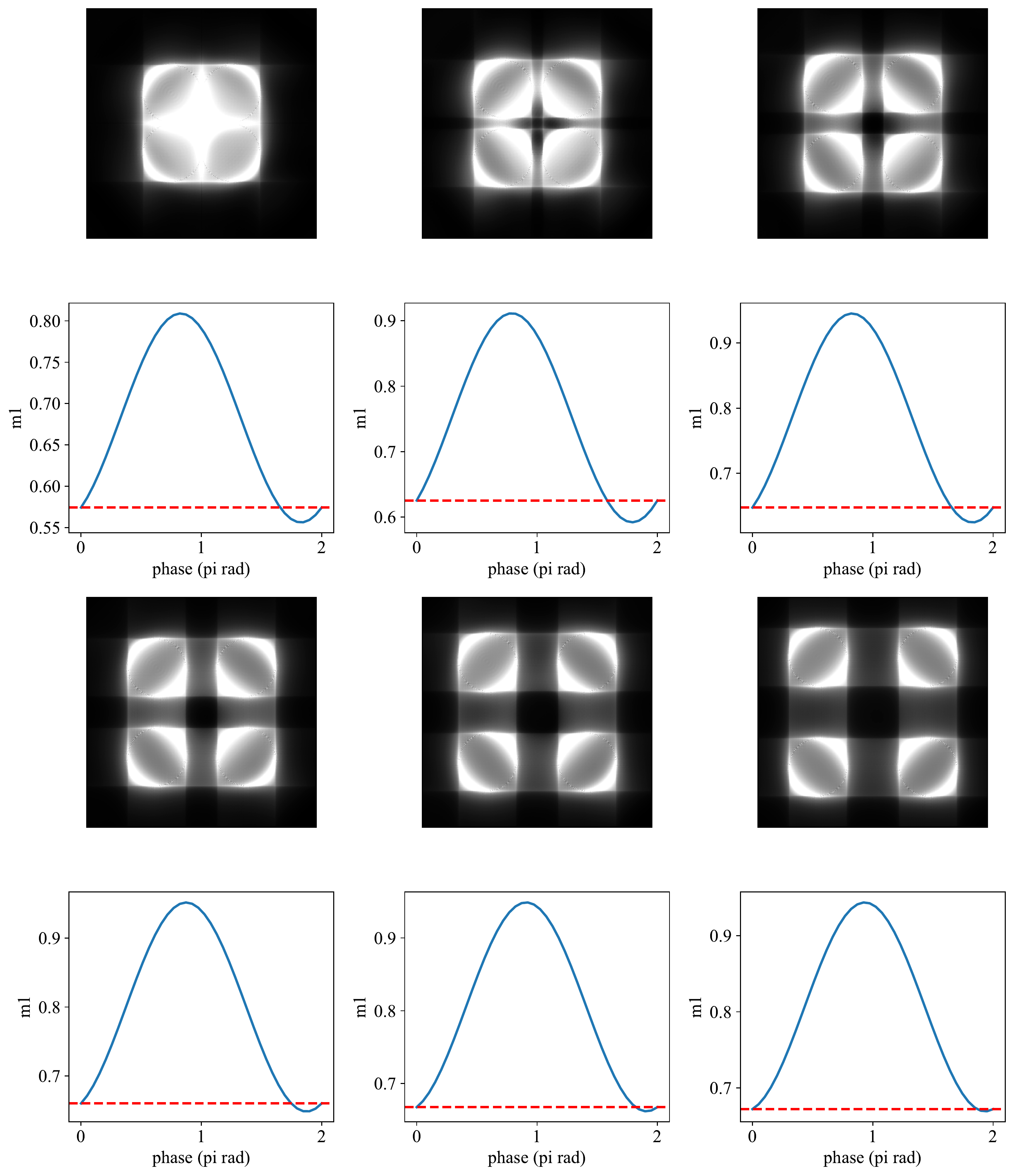}
    \caption{Dependence of PWFS tilt angle on m1 vs. additive piston phase curve (as in Fig. \ref{fig: bpwfs}b). The pupil intensity image for the additive piston phase at which m1 is minimized is shown above each corresponding curve. All images are displayed on the same relative contrast scale.}
    \label{fig: pup_sep}
\end{figure}

Fig. \ref{fig: bpwfs} illustrates the bPWFS and dPWFS concepts.
As shown, the bright and dark PWFSs are different additive piston phase offsets to the regular PWFS tilt angles, such that
\begin{align}
    \label{eq: bPWFS}\phi_\mathrm{bPWFS}&=\phi_\mathrm{PWFS}+\phi_\mathrm{b},\;\mathrm{with} \\
    \phi_\mathrm{b}&=1.7\pi\;[\mathrm{rad}]\;\forall\; r\le0.5\;\lambda/D,\;\mathrm{and} \nonumber \\
    \phi_\mathrm{b}&=0\;[\mathrm{rad}]\;\forall\; r>0.5\;\lambda/D, \nonumber
\end{align}
where $\phi_\mathrm{PWFS}$ is the regular PWFS focal plane phase shift tilt angles, $\phi_\mathrm{b}$ is the additive phase component for the bPWFS, and $r$ is a radial polar grid in natural units of $\lambda/D$. In general the complex electric field at the focal plane of the bPWFS optic is represented by $A_\mathrm{sys}\;e^{i(\phi_\mathrm{sys}+\phi_\mathrm{bPWFS})}$, where $|A_\mathrm{sys}\;e^{i\;\phi_\mathrm{sys}}|^2$ is the system point spread function (PSF) if imaged on a dector placed in that focal plane. For the dPWFS, the analogous $\phi_\mathrm{dPWFS}$ term is identical to equation \ref{eq: bPWFS} other than $\phi_\mathrm{d}=0.8\pi\;[\mathrm{rad}]\;\forall\; r\le0.5\;\lambda/D$. As shown in Fig. \ref{fig: bpwfs}b, the value of this piston offset produces drastically different pupil images compared to the regular PWFS, some with more light diffracted into the pyramid pupil footprints (the bPWFS) and some with less light diffracted into the footprints (the dPWFS). Fig. \ref{fig: pup_sep} then shows how this diffractive modification to the PWFS pupil images changes with the tilt angle defining the pupil separations, 
illustrating that increased pupil ``throughput'' (i.e., a decreased value of m1) with this bright PWFS concept relative to the regular PWFS is only achieved when the pupil separations are less than about 1 pupil diameter from one another. This effect is to be expected; effectively, this additive phase shift is acting as a Roddier\&Roddier coronagraph\cite{r_and_r} or a Zernike WFS\cite{zernike} on the four pupils, depending on the value of additive piston applied to the central $\lambda/D$ diameter region. When the four pupils are separated by more than one pupil diameter form one another, the central dark ``coronagraphic'' pupil from the bPWFS no longer redistributes light from within this central pupil footprint into the four pupils from the PWFS tilt angles. However, a more complex numerical optimization should further investigate the limits of this approach (see \S\ref{sec: conclusion}). Regardless, a separation between the four PWFS pupils is still generally a desired trait, enabling (1) an increased achievable pupil sampling for use with higher order deformable mirrors (DMs) and/or to reduce aliasing in WFS measurement errors, and/or (2) decreasing the required sub-array/detector size to readout, enabling higher accessible loop speeds and/or lower detector read noise.
\subsection{Error Budget Analysis}
\label{sec: error_budget}
Instead of considering a full end-to-end closed-loop error budget analysis, in this paper we only implement a differential analysis to compare performance between the regular PWFS and bPWFS (both without tip/tilt modulation). Although this approach cannot be used to predict achievable closed-loop Strehl ratios, it gives insight into how a bPWFS could improve closed-loop wavefront errors over the non-modulated PWFS (i.e., using modulation to close the loop initially, but then subsequently not using modulation; also see \S\ref{sec: modulation}).

The setup of our differential error budget analysis is identical to Ref. \citenum{ben} (Sec. 2.3) and also briefly summarized here. We first describe our simulation and calibration assumptions. We use a 1024$\times$1024 pixel image with a 191 pixel diameter to simulate the entrance pupil wavefront, assuming a square DM with 32 actuators across the pupil diameter. We use the same tilt angle/pupil separation as illustrated in Fig. \ref{fig: bpwfs}. A monochromatic wavelength of 1.65 $\mu$m is assumed to convert between radians and nm of wavefront error. Input residual atmospheric phase screens are generated assuming a -2 power law and random phase, normalized to 100 nm root mean square (rms) within the DM control region. An 8m diameter telescope is assumed with no secondary obscuration or spiders for simplicity. Assumed transmission through the atmosphere, instrument throughput, and detector quantum efficiency are 90\%, 80\%, and 80\%, respectively. A 1\% bandpass is assumed for flux calculation purposes with an otherwise monochromatic Fraunhofer propagation simulation (see \S\ref{sec: chromaticity}). To generate a command matrix, we use a classical zonal control approach, assuming perfect uncoupled Gaussian DM actuator influence functions whose full width at half maximum (FWHM) are one actuator pitch. The pixel values within the four pupil footprints in the PWFS/bPWFS detector plane for a push minus pull of each DM actuator are used to generate an interaction matrix. The singular value decomposition cutoff of the matrix inversion is separately optimized for both the PWFS and bPWFS to enable the lowest total residual wavefront error after one iteration on the above-described input 100 nm rms phase screen. An input target image to be multiplied by the command matrix to generate DM commands is a differential measurement, in this case the difference between the target WFS image and the diffraction-limited WFS image for the corresponding Fourier-based WFS phase mask.

\begin{figure}[!h]
    \centering
    \begin{subfigure}[b]{0.54\textwidth}
        \includegraphics[width=\textwidth]{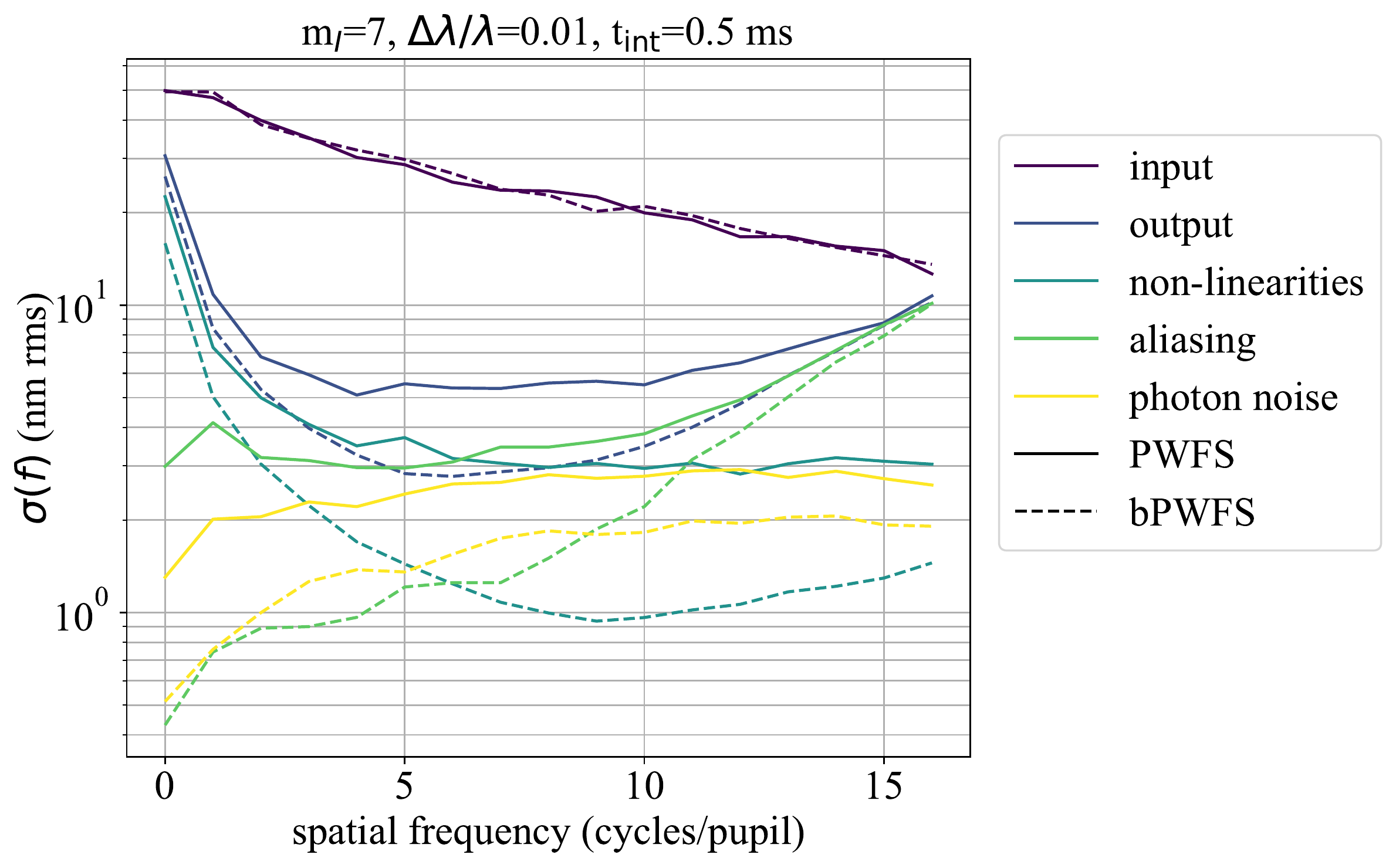}
        \caption{}
    \end{subfigure}
    \hfill
    \begin{subfigure}[b]{0.44\textwidth}
        \includegraphics[width=\textwidth]{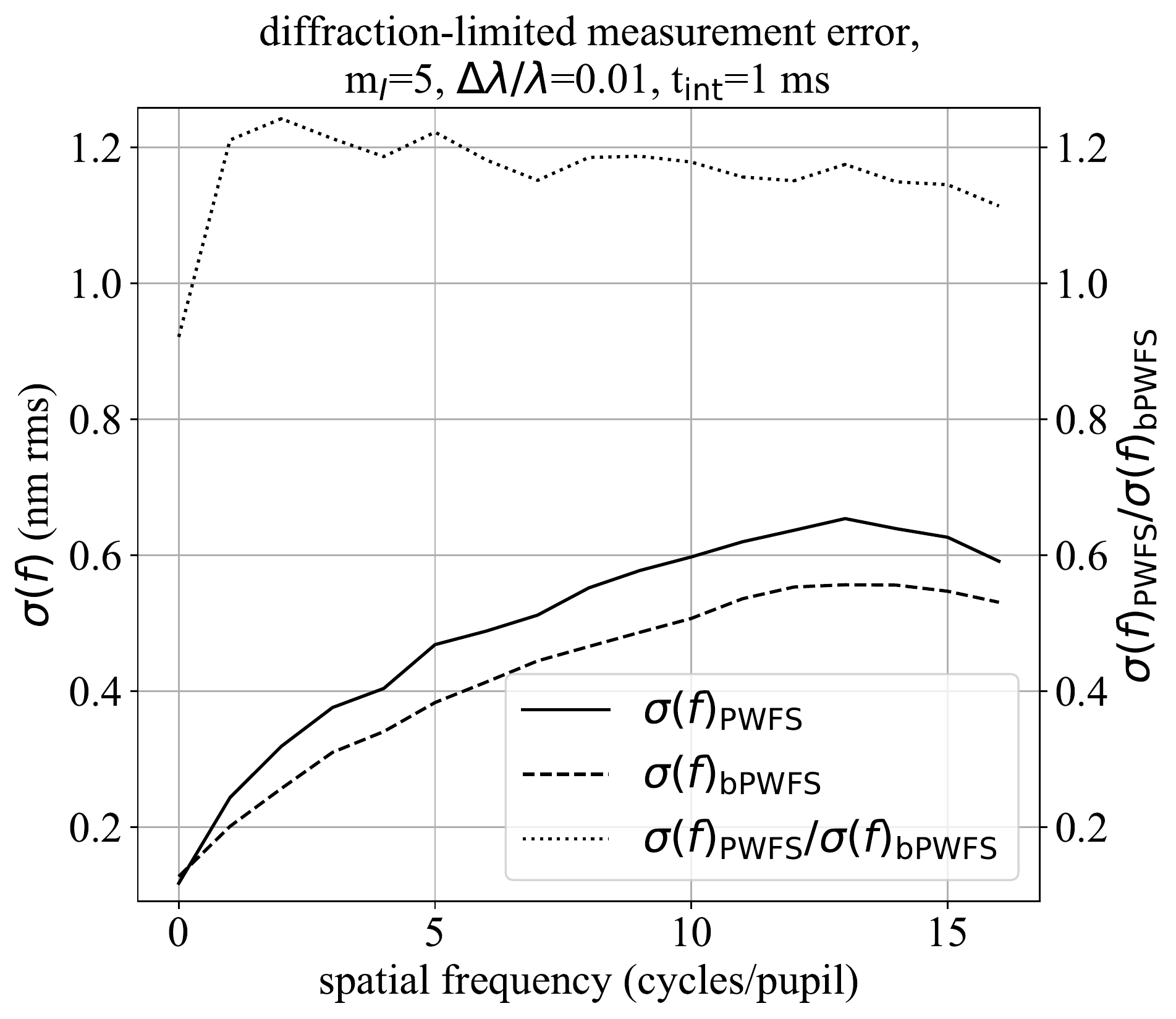}
        \caption{}
    \end{subfigure}
    \caption{AO error budget, comparing the regular PWFS (solid lines) and bright PWFS (dashed lines). In both panels the y axis shows wavefront error for a given spatial frequency bin, identical to the definition in equation 1 of Ref. \citenum{zelda}. (a) A breakdown of output residuals, after 3 iterations using a unity gain integral controller, into components for photon noise, alaising, and non-linearities, comparing the PWFS and bPWFS for each component. Each curve shown is from a median over 100 different uncorrelated wavefront realizations (where in each case the input is normalized to 100 nm rms as described in the text). (b) Error budget comparison of reconstructed photon noise for the PWFS and bPWFS, both for an input diffraction-limited (i.e., zero phase) wavefront. Each curve is medianed over 100 individual photon noise realizations.}
    \label{fig: error_budget}
\end{figure}

Fig. \ref{fig: error_budget} shows the results of our differential error budget analysis, comparing the PWFS and bPWFS. 
The three non-linearities, aliasing, and photon noise curves are calculated as follows (again only briefly summarized here; see Ref. \citenum{ben} for a full description).
\begin{enumerate}
    \item Non-linearities are defined as the residual wavefront when no photon noise is included in the simulation and an algorithmic anti-aliasing filter on the entrance pupil wavefront error is applied.
    \item The aliasing term is the differential measurement between the residual wavefront without an anti-aliasing filter (aliasing + non-linearities) and with an anti-aliasing filter (non-linearities), resulting in only the aliasing component (in both cases without including simulated photon noise).
    \item The photon noise term is the differential wavefront between the residual wavefront with photon noise simulated (photon noise + non-linearities) and with no photon noise simulated (non-linearities), in both cases with an anti-aliasing filter applied.
\end{enumerate}

Figure \ref{fig: error_budget} clearly shows the gain in all error budget terms from the bPWFS over the PWFS: the overall residual wavefront in panel a is lower, with relatively less contributions from non-linearities, aliasing, and photon noise. Panel b shows that for the same number of input photons, the measurement error due to photon noise for a diffraction-limited wavefront (i.e., no aliasing and minimal non-linearities) is about 20\% lower for the bPWFS vs. PWFS. These performance gains can be interpreted as follows.
\begin{enumerate}
    \item Non-linearities and aliasing. The bPWFS phase mask diffracts more light into the four pupil footprints and enables that increased light to display entrance pupil phase aberations as intensity variations on the WFS detector, which causes less signal from the input wavefront error to be spatially redistributed beyond such footprints. More diffracted light outside the pupil footprints leads to increased non-linearities, particularly for larger wavefront errors. More diffracted light/light redistribution also has the potential to optically redistribute higher order un-correctable wavefront errors to look like correctable ones and vice versa.
    \item Photon noise. More photons within the pupil footprints that show signal from entrance pupil phase aberration as intensity variations increases the signal-to-noise ratio (S/N) of a given wavefront error proportional to the square root of the number of photons. A higher S/N WFS measurement of a given wavefront error results in less noise being propagated onto the DM from the WFS reconstruction and ultimately a lower achievable residual wavefront error.
\end{enumerate}
Also note that this same error budget analysis for the dPWFS showed the opposite of what we presented here for the bPWFS: all error terms for the dPWFS perform worse than the PWFS. This is consistent with the above interpretation, illustrating that when more light is diffracted outside the pupil footprints which would otherwise be entrance pupil phase aberrations converted to intensity variations on the WFS detector, expected performance is worse.
\section{LABORATORY TESTS}
\label{sec: lab}
\subsection{The LOOPS Testbed}
\label{sec: loops_intro}
\begin{figure}[b]
    \centering
    \includegraphics[width=0.95\textwidth]{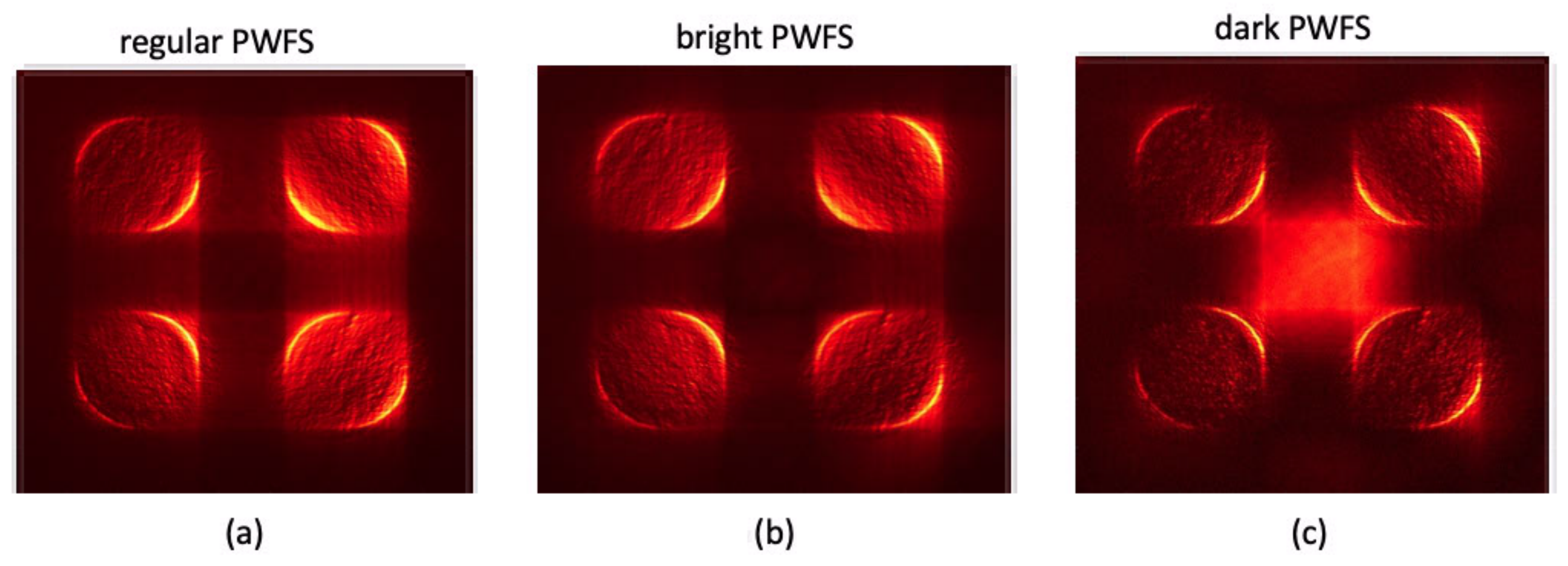}
    \caption{Preliminary Laboratory validation of the bright PWFS concept with the LOOPS testbed at LAM, using an SLM in the focal plane to modulate the added focal plane piston phase shift within the central 0.5 $\lambda/D$ of the optical axis. (a) The regular PWFS (i.e., no added piston). (b) the bright PWFS: 1.7 $\pi$ rad added piston. (c) the dark PWFS: 0.6 $\pi$ added piston. }
    \label{fig: lab}
\end{figure}
The LOOPS---LAM (Laboratoire d’Astrophysique de Marseille)/ONERA (Office National d’{\'E}tudes et de Recherches en A{\'e}rospatiales) On-sky Pyramid Sensor---testbed is a laboratory at LAM developed to test new AO technologies, described in Ref. \citenum{loops_testbed} and references therein. Although many new AO technologies have been deployed and tested on this bench, for the purposes of this paper there are three relevant hardware sources on the LOOPS bench (introduced in the order of light propagation): (1) a $\lambda$=635nm laser diode light source conjugated at infinity, (2) a downstream 1024$\times$1280 pixel Hamamatsu Spatial Light Modulator (SLM; LCOS-SLM X13138) conjugated to a focal plane at f/264\footnote{See Ref. \citenum{loops_testbed} for a detailed description of the steps taken to limit SLM-based systematics (such as polarization-dependent and filling-factor errors), ensuring custom focal plane phase shifts can be applied without interference from these effects.}, and (3) a further downstream a 2048$\times$2048 pixel Hamamatsu ORCA-Flash 4.0 v2 pupil imaging detector, sampling around 300 pixels across the pupil diameter.
\subsection{Preliminary Results}
\label{sec: lab_results}
The LOOPS focal plane SLM described in \S\ref{sec: loops_intro} enables applying the phase masks for the PWFS, bPWFS, and dPWFS, with all other testbed parameters unchanged between masks. The results of this are shown in Fig. \ref{fig: lab}.
Consistent with simulations in Fig. \ref{fig: bpwfs}a, Fig. \ref{fig: lab} clearly shows less light between the four pupils for the bPWFS vs. regular PWFS and conversely more light between pupils for the dPWFS.
\section{DISCUSSION AND FUTURE WORK}
\label{sec: discussion}
\subsection{Modulation and Linearity}
\label{sec: modulation}
Implementing the modulated PWFS is a key next step to validating the benefit of the bPWFS over the PWFS. We expect that for modulation radii greater than about 1 $\lambda/D$, the modulated bPWFS performance will be identical to the modulated PWFS, since the bPWFS piston mask only extents radially to 0.5 $\lambda/D$. However, a key question to answer is whether or not switching from a modulated to un-modulated bPWFS in closed-loop will enable improved closed-loop performance. 
Future simulations and/or laboratory tests should therefore investigate this further. 

Alternatively, another approach to considering the benefit of the bPWFS is it's use as a second stage AO WFS (i.e., both a first stage AO WFS, such as a modulated PWFS or a Shack Hartmann WFS, and a second stage bPWFS control the same common path DM). In this configuration, although a more optically complex system with multiple WFSs, we have essentially already demonstrated the benefit of this approach in Fig. \ref{fig: error_budget}a, showing that for input 100 nm rms realizations (i.e., ExAO-level first stage wavefront residuals) the bPWFS consistently enables better achievable residual wavefront errors after a second stage correction than the PWFS. However, to be fair this comparison does not include other Fourier-based WFSs that require a diffraction-limited input wavefront, such as the Zernike\cite{vincent_zwfs} or iQuad\cite{iquad} WFS. Also note that a first and second stage WFS both controlling a common DM requires a more advanced control framework to ensure that any differential NCPA and/or chromaticity (if the two WFSs are operating with different bandpasses) doesn't cause instability; Ref. \citenum{ben} has investigated this problem and published preliminary solutions that would define non-conflicting spatio-temporal control authorities for each WFS.
\subsection{Chromaticity and Fabrication}
\label{sec: chromaticity}
Two forms of wavelength-dependent wavefront evolution (i.e., chromaticity) must be addressed in order to enable this technique operational over a reasonable bandpass: (1) PSF magnification with wavelength on the bPWFS focal plane phase mask and (2) scalar piston evolution with wavelength. We address each point below.
\begin{enumerate}
    \item PSF magnification. In principle, a Wynne corrector\cite{wynne} could remove this effect, although it would then cause a corresponding downstream pupil magnification with wavelength, which could then be mitigated by a second ``inverse'' Wynne corrector. However, PSF magnification at 0.5 $\lambda/D$ radius is a weak effect: a $\Delta\lambda/\lambda\sim60\%$ bandpass from 500-900 nm would move the piston masks' radial edge between 0.65 and 0.3 $\lambda/D$ at 500 and 900 nm, respectively. This is a minimal change in radius that would likely not significantly modify the bPWFS's diffractive effect, but should be invenstigated further in simulation to confirm.
    \item Scalar piston phase chromaticity. If the bPWFS were fabricated as a scalar mask (e.g., via etching, metal deposition, and/or other similar depth-dependent methods), the applied piston offset in radians would be chromatic: ignoring PSF magnification effects, for the same 60\% bandpass from the previous point a 1.7$\pi$ rad phase shift at 700 nm would vary between 1.2$\pi$ and $2.2\pi=0.2\pi$ rad at 500 and 900 nm, respectively. Fig. \ref{fig: bpwfs}b already shows that this effect would completely un-do the bPWFS benifits, on average over the full bandpass causing more light to be diffracted outside the pupil footprints compared to the regular PWFS. Therefore, an achromatic phase piston shift must be applied instead, likely requiring a ``two stage'' PWFS system (i.e., one intermediate focal plane for the achromatic piston phase mask and second separate intermediate focal plane the pyramid phase mask). At least two phase mask technologies exist that could be used to fabricate the above-described achromatic bPWFS piston mask: sub-wavelength gratings\cite{subwavelength} and liquid crystals\cite{liquidcrystals}.
\end{enumerate}
\section{CONCLUSION}
\label{sec: conclusion}
In this paper we have introduced a new modification to the Pyramid wavefront sensor (PWFS) phase mask called the bright PWFS (bPWFS). This bPWFS phase mask, which simply adds a 1.7$\pi$ radian piston phase shift within a 0.5 $\lambda/D$ radius of the optical axis on top of the existing PWFS tilt angles, leverages the diffractive nature of a Fourier-based WFS to increase the total amount of light collected within the four PWFS pupil footprints, described further in \S\ref{sec: concept}. Using this design, we showed the following.
\begin{enumerate}
    \item A simulated error budget analysis in \S\ref{sec: error_budget} illustrated that the bPWFS compared to the PWFS enables a residual wavefront error with relatively lower impact from non-linearities, aliasing, and photon noise, all of which can be attributed to the boosted cumulative intensity within the WFS pupil footprints.
    \item In \S\ref{sec: lab_results} we validate the bPWFS concept on the LOOPS testbed at LAM, which includes a focal plane spatial light modulator to enable a comparison with the same system between the PWFS and bPWFS. Recorded PWFS and bPWFS pupil images confirm that the bPWFS increases cumulative intensity within the pupil footprints, consistent with simulations.
\end{enumerate}
Lastly, in \S\ref{sec: discussion} we outlined further steps to consider in deploying this technology into real AO systems, including use of a modulation tip/tilt mirror and/or second stage WFS (\S\ref{sec: modulation}) and addressing potential issues of chromaticity and possible optical and fabrication solutions (\S\ref{sec: chromaticity}). 

More generally, the simple additive piston phase mask design presented here for the bPWFS is by no means fully-optimized. Numerical coronagraph design optimization is a mature field of research\cite{coronagraph_design}; similar techniques should be explored for Fourier-based pupil plane WFS design (e.g., considering options for complex phase masks, multiple pupil and/or focal planes, etc., analagous to current numerical approaches to coronagraph design), with the difference that measurement errors, non-linearities, aliasing, and other error budget terms be optimally minimized (i.e., in contrast to minimizing stellar throughput while maximizing exoplanet throughput for coronagraph design techniques).
\section*{Acknowledgments}
We gratefully acknowledge research support of the University of California Observatories for funding this research. We thank Beno{\^i}t Neichel, Dominic Sanchez, Phil Hinz, Thiery Fusco, Maaike van Kooten, and Jules Fowler for comments, suggestions, and discussions that have contributed to this manuscript.

\bibliography{report} 
\bibliographystyle{spiebib} 

\end{document}